\makeatletter

\newenvironment{inlinefigure}{%
\def\@captype{figure}%
\noindent\begin{minipage}{0.999\linewidth}\begin{center}}
{\end{center}\end{minipage}\smallskip}
\makeatother

\documentclass[10pt,preprint]{aastex}
\usepackage{epsfig}
\usepackage{emulateapj5}

\newcommand{\mdot}{\dot{M}}

\newcommand{\msun}{{M}_{\odot}}
\newcommand{\rsun}{{R}_{\odot}}
\newcommand{\msyr}{\msun \ {\rm yr^{-1}}}

\slugcomment{To appear in the Astrophysical Journal}

\shorttitle{Disk-Jet Connection in CH Cygni}
\shortauthors{Sokoloski \& Kenyon}

\begin{document}


\title{CH Cygni I: Observational Evidence for a Disk-Jet Connection}


\author{J. L. Sokoloski and S. J. Kenyon}
\affil{Smithsonian Astrophysical Observatory, 60 Garden St., Cambridge, MA 02138}
\email{jsokoloski@cfa.harvard.edu}


\begin{abstract}
We investigate the role of accretion in the production of jets in the
symbiotic star CH Cygni.  Assuming that the rapid stochastic optical
variations in CH Cygni come from the accretion disk, as in cataclysmic
variables, we use changes in this flickering to diagnose the state of
the disk in 1997. At that time, CH Cyg dropped to a very low optical
state, and Karovska et al. report that a radio jet was produced.  For
approximately one year after the jet production, the amplitude of the
fastest (time scales of minutes) variations was significantly reduced,
although smooth, hour-time-scale variations were still present.  This
light curve evolution indicates that the inner disk may have been
disrupted, or emission from this region suppressed, in association
with the mass-ejection event.  We describe optical spectra which
support this interpretation of the flickering changes. The
simultaneous state change, jet ejection, and disk disruption suggests
a comparison between CH Cygni and some black-hole-candidate X-ray
binaries that show changes in the inner disk radius in conjunction
with discrete ejection events on a wide range of time scales (e.g.,
the microquasar GRS 1915+105 and XTE J1550-564).
\end{abstract}




\section{Introduction}

Many astrophysical systems produce collimated jets, including young
stellar objects, X-ray binaries and symbiotic stars, and active
galactic nuclei.  \cite{liv97} suggested that there is a common
formation mechanism for all jets, and that they are all fundamentally
accretion-powered.  If these assertions are correct, then accreting
white-dwarf (WD) systems play a vital role in the study of jet
phenomena, since WD disks are much more well understood than, for
instance, black-hole disks (see, e.g., Warner 1995).  Furthermore,
\cite{zammar02} speculated that the relationship between WD and
black-hole jet sources could be significant enough that some symbiotic
stars, including CH Cygni, can be considered ``nanoquasars.''

In stellar black-hole systems, 
the presence of bi-polar radio jets is related to the state of the
accretion disk.  Outflows may be present in the low/hard X-ray state,
but not in the high/soft state
\citep[e.g.,][]{hh95,fen01}. 
Transitions between 
states can also lead to discrete plasma ejections
\citep[][]{hh95,kuul99,fen99a,fen99b,fk01}. 
Collimated jets, however,
have not been observed in cataclysmic variables
\citep[CVs;][]{kl98}, the most 
numerous and well-studied type of WD interacting binary.  Although WD
systems provide the best details on the structure of the disk, they
have so far told us little about jets.

CH Cygni, on the other hand, is one of the most dramatic Galactic jet
sources.  At least three distinct sets of jet events have been
recorded, in 1984/85, in 1994/95, and in 1996/97
\citep{tay86,kar98,crock01}.  Elongated radio structure from each of these
ejections was detectable for several years
\citep[e.g., in 1986, 1995, and 1999][]{crock01}, and
each jet event was associated with a period of optical activity.  For
all three events, jet production followed a sudden drop in the optical
flux.

In this paper, we investigate the role of accretion in the production
of jets 
in CH Cygni.  The optical spectrum of CH Cyg is complicated by the
presence of the red giant and nebular emission; examining rapid
(time scales of minutes to hours) optical variations is one way to
isolate emission from the accretion disk.
Although the exact nature of stochastic optical variations observed in
non-magnetic CVs is not clear, they are generally
considered the result of accretion onto a WD through a disk
\citep[e.g.,][]{war95}. Proposed mechanisms for disk flickering
include unstable mass transfer from the mass-donor star, magnetic
discharges, turbulence, and boundary layer instabilities (Bruch 1992).
If a disk can form from the red-giant wind\footnote{See \cite{sk03}
(hereafter paper II) for a 
discussion of this point.}, 
the flickering in CH Cygni probably has a similar origin.

Between 1997 and 2000, we performed rapid optical photometric and
optical spectroscopic observations of CH Cygni.  We describe the
complete data set in a companion paper (Sokoloski \& Kenyon 2003;
hereafter paper II).  In the present paper, we discuss observations
from 1997, which provide information about the accretion disk
during the production of a radio jet.

The paper is divided into six sections.  After the initial
description of the observations and fast-photometry results in
\S\ref{sec:obs}, we discuss the 1997 radio jet in
\S\ref{sec:97jet}.   In \S\ref{sec:ddisrupt}, we interpret the flickering
changes seen when a radio jet was produced 
as the result of disruption of the inner accretion disk, or
suppression of the emission from this region.  Several alternative
interpretations are examined in \S\ref{sec:alt}. We explore the
implications of these results for jet formation in CH Cygni, and by
extension other white-dwarf accretors in \S\ref{sec:implications}.

\section{Observations and Photometric Results} \label{sec:obs}

We performed high-time-resolution optical differential photometry at
$B$ and at $U$, using the 1-meter Nickel telescope at Lick Observatory
between 1997 and 2000.
In this paper, we focus on observations from around the time of the
1997 radio jet.  In addition, P. Berlind, M. Calkins, and several
other observers acquired low-resolution optical spectra of CH Cygni
with FAST, a high throughput, slit spectrograph mounted at the Fred
L. Whipple Observatory 1.5-m telescope on Mount Hopkins, Arizona
\citep{fab98}.  See Sokoloski, Bildsten, \& Ho 2001 for description of
photometric observing technique and instruments, and paper II for
details of the CH Cyg data analysis.

\begin{inlinefigure}
\plotone{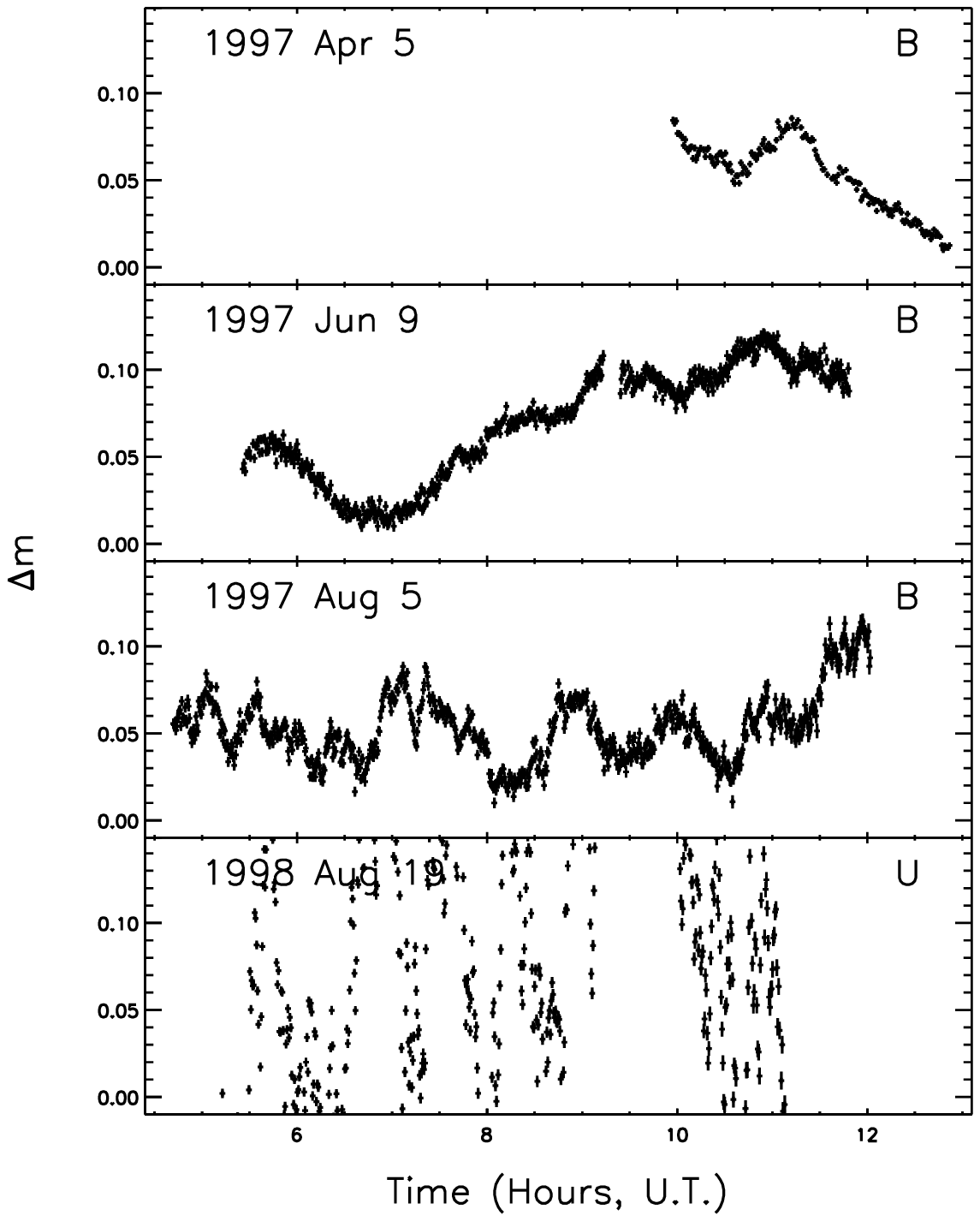} 
\caption{
Example light curves from 1997 and 1998 (from
paper II), plotted on the same scale for direct comparison.  The light
curves evolve from showing smooth, low-amplitude variations after the
ejection of the radio jet in 1996 (see \S\ref{sec:97jet}), to full
CV-like flickering more than one year later (the amplitude of the
variations in 1998 span three times the magnitude range shown).\label{fig:alllcs}}
\end{inlinefigure}

Example high-time-resolution light curves from 1997 and 1998 are shown
in Figure~\ref{fig:alllcs}.  The first two observations, from 1997
April and June, show low-amplitude ($\Delta m < 0.15$ mag), smooth
variations, and have power spectra that generally cannot be fit with
power-law models (similar light curves were reported by \cite{rod97}
as early as 1996 June).
A few months later, in 1997 August, the 
fractional flickering amplitude had not changed, but the strength of
the fastest (minute-time-scale) variations increased.  The light curve
then had a more jagged appearance.  The power spectrum of this
observation had the standard power-law shape found in non-magnetic
CVs.  One year later, in 1998 July and August, the variability
amplitude increased to the typical high-state value of $\Delta m \sim
0.5$ mag while the power spectrum retained its power-law shape.
Figure~\ref{fig:pdss} compares a power spectrum from the post-jet
period when only smooth variations were seen with a power spectrum
from a few months later, when minute-time-scale variations returned.

\begin{inlinefigure}
\plotone{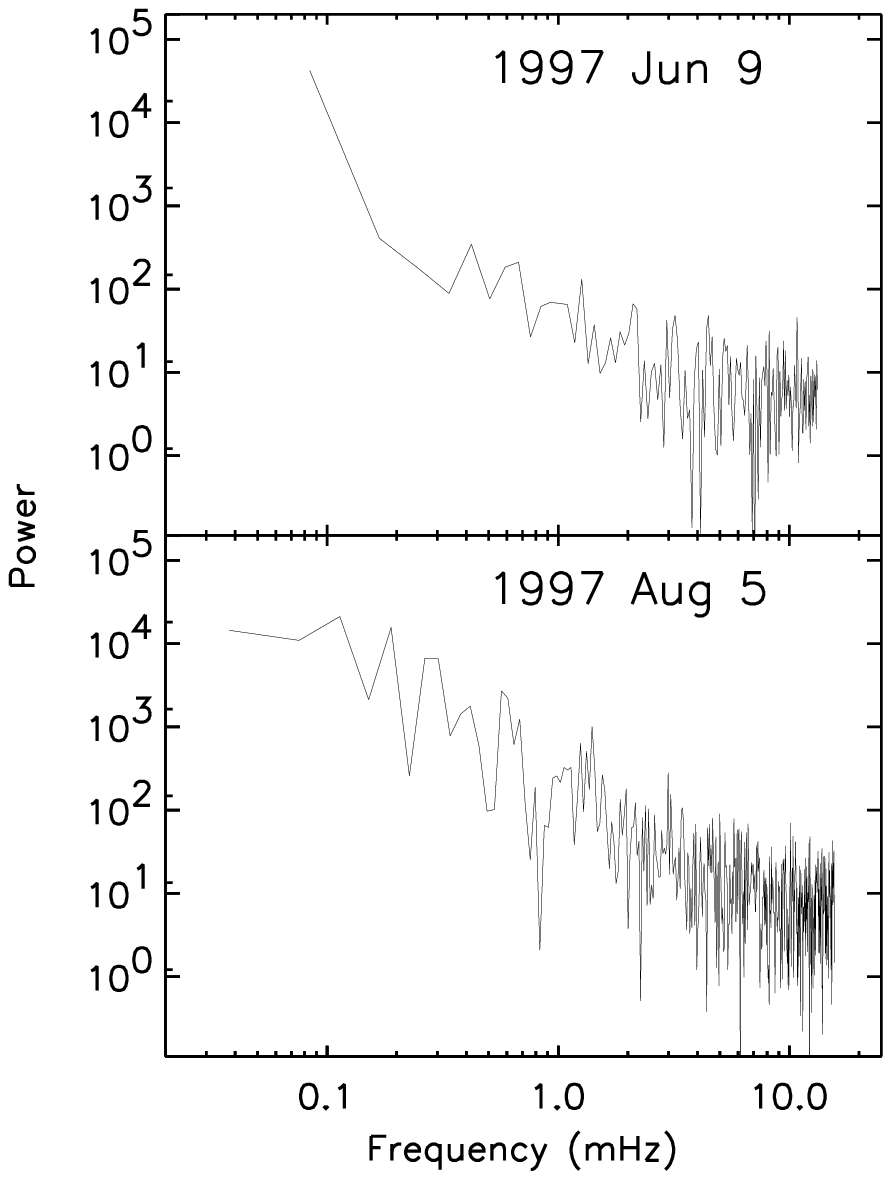} 
\caption{
Example power spectra (normalized by the total number of source counts
in each observation) from 1997.  The power spectrum during the early
production of the 1997 radio jet (top) cannot be fit with a simple
power-law model.  A power spectrum from several months later (bottom
panel) is well described by a power-law model.
\label{fig:pdss}}
\end{inlinefigure}

\section{The 1997-1999 Radio Jet} \label{sec:97jet}

Based on the radio observations reported by
\cite{kar98},
CH Cygni began producing a radio jet sometime before 1997 January,
when they first detected extended radio emission.  The radio flux
density began to rise in 1997 April, at the same time as we started
our photometric observing campaign.  Radio extension associated with
the radio brightening was later confirmed by Karovska (private
communication);
\cite{eyres02} discuss the mass outflow from CH Cygni during and after
1998.  
Figure~\ref{fig:jet} shows the long-term optical light curve of CH
Cygni (kindly provided by the AAVSO), with the times of our fast
photometric observations marked, and the early development of the radio
jet at 22 GHz \citep{kar98}.

\setcounter{figure}{2}
\begin{figure*}
\epsfig{file=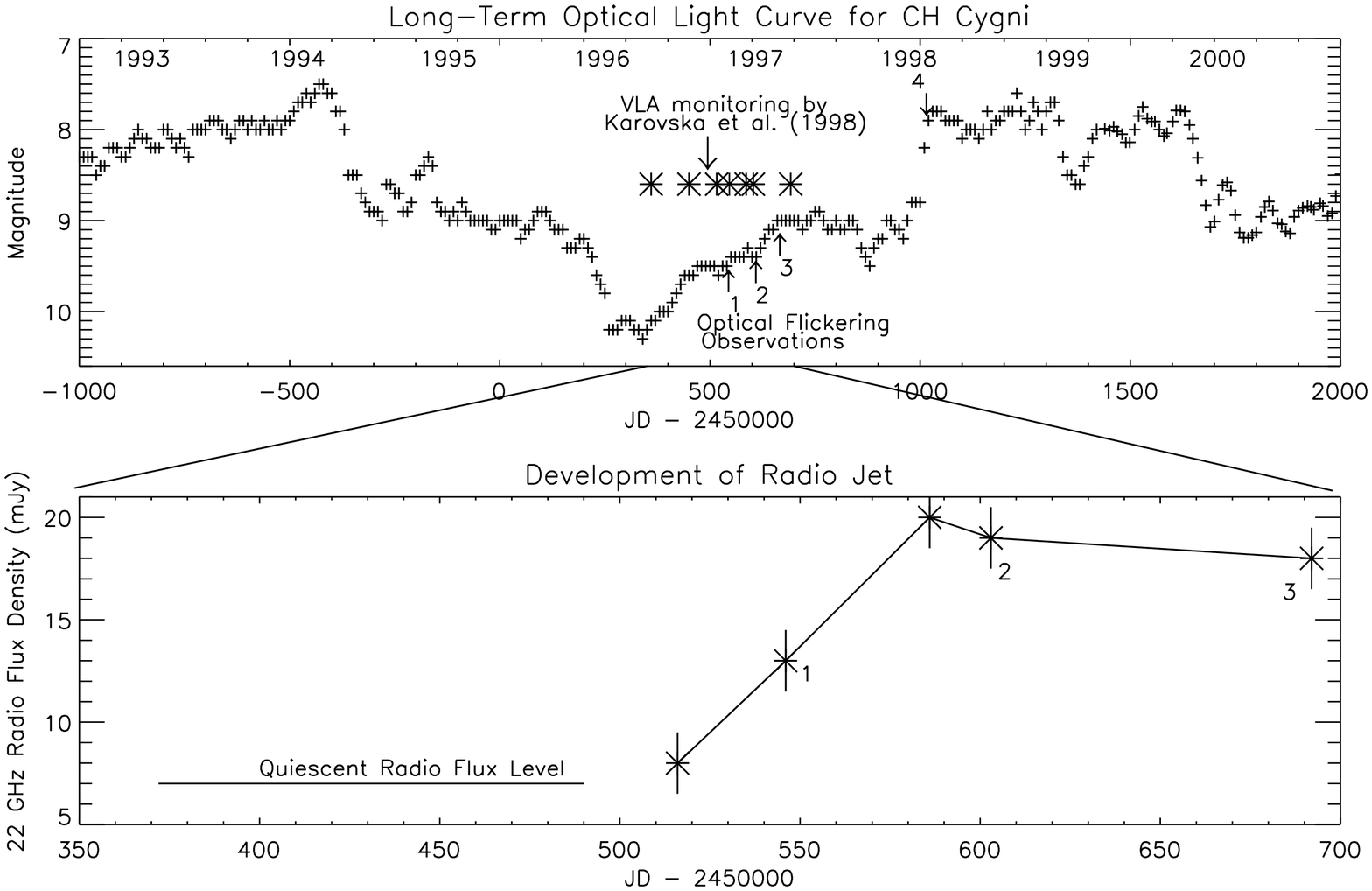,width=6.5in,height=3.5in}
\caption{\footnotesize Top panel: long-term optical light curve of CH Cygni,
from the AAVSO.  Four flickering observations in 1997 and 1998 are
marked with arrows, and the period of radio monitoring by Karovska et
al. (1998) is marked with stars. Bottom panel: 22 GHz radio flux
densities, from Karovska et al. (1998).  Detection of spatial
extension in a follow-up VLA observation confirmed that the rise in
the radio emission was indeed due to the production of a jet
(Karovska, private communication).
\label{fig:jet}}
\end{figure*}

We can estimate the speed of the jet material, and hence the
approximate time when the material was expelled, if we assume that the
radio elongation measured in 1999 by \cite{crock01} was from the 1997
ejection.  The elongation measured by \cite{crock01} was primarily
south of the central radio source.  Either the ejection was one-sided,
or the emission from a northern component had faded by 1999 September.
Comparing the initial report of north-south elongation at 0.25\arcsec
resolution in 1997 January
\citep[at 15 GHz with the VLA, in BnA configuration][]{kar98} with the
approximately 1.7\arcsec\, north-south structure detected by
\cite{crock01} in 1999 September (at 8.4 GHz with VLA in A
configuration), we estimate a growth rate of roughly 1.5\arcsec\, in
990 d, or 650 km s$^{-1}$ for a distance of 245$\pm 50$ parsec (from
the mode of the Hipparcos parallax probability distribution).  In
making this estimate, we took the southerly extension to be
approximately 0.2\arcsec\ in 1997 Jan, and measured the 1999 jet size
between the central radio component and the maximum extent of the
$3\sigma$ contour \citep[from the map of][]{crock01}.  At that rate,
the jet took roughly 130 d to expand the initial 0.2\arcsec, putting
the initial expulsion of material around JD 2450330 (1996
August/September).  This date for the initial ejection is only
approximate, since we do not have a map of the radio emission in 1997,
we do not know how the sensitivities of the 1997 and 1999 observations
compare, and the 1997 and 1999 observations were done at different
frequencies.  Even so, our estimate puts the origin of the jet roughly
within a few months of the 1996 optical fading and the beginning of
the period where only smooth variations were apparent in the optical
light curves (as in the top panels in Figure~\ref{fig:alllcs}).  It is
therefore likely that the flux drop, the jet ejection, and the change
in the optical flickering were all related.

In CH Cygni, the production of jets generally appears to follow a
sudden drop in the optical flux.  In 1984/85, the 14.9 GHz radio flux
density increased by a factor 30 as the initial radio structure grew
to 0.4\arcsec
\citep{tay86}.  The production of this jet followed a decrease in the
visual flux of over 1.2 mag in less than 50 days (AAVSO).  The 1995
radio jet may have been related to a similar optical decline in 1994
\citep{crock01},
and the 1997 jet followed a 1 mag drop in
about 100 d (AAVSO).  
In 1986,
\cite{crock01} found that the core of the radio emission had a
positive spectral slope, whereas the extended regions had a negative
spectral slope consistent with non-thermal synchrotron emission from
relativistic electrons in a magnetic field.  They suggested that the
electrons were accelerated in a shock as the ejecta collided with
nebular material from the circumstellar wind.  In 1999, they found
similar negative-spectral-index emission from the extended region
presumably associated with the 1997 jet ejection.

Although there are some similarities between the ejections in 1984/85
and 1997, there are also differences.  In 1984/85, the 14.9 GHz radio
flux density peaked roughly 6 months after the optical drop.  The
delay between the optical decline in 1996 and the subsequent radio
rise was approximately 1 yr.  Correspondingly, the expansion velocity
estimated for the 1985 jet of approximately 1300 km s$^{-1}$ (using
the same method of comparing the distance between the central radio
component and the farthest $3\sigma$ contour, and taking $d=245$
parsec) is roughly twice our estimate of 650 km s$^{-1}$ for the 1997
jet (see above).  Moreover, in 1997, the optical flux decreased from
an already low state to a much lower level.  So either the system
brightness was affected by dust obscuration at this time, there was an
intrinsic fading of the red giant \citep[as suggested by][]{sko97}, or
the 1996 fading put the disk into a new, very low state.

\vspace{-0.3cm}
\section{Disk Disruption} \label{sec:ddisrupt}

In almost every type of interacting binary star (or other accreting
system), disk accretion produces stochastic brightness
variations.  
Assuming that the rapid stochastic variations in CH Cygni are due to
disk accretion\footnote{See paper II for a detailed discussion of this point.},
changes in the flickering can tell us about changes in the
disk.
Although the physical mechanism responsible for disk flickering is not
well understood, several possible models suggest that the fastest
variations come from the innermost regions of the disk.  This radial
dependence arises because both the viscous and dynamical times
increase with disk radius.  For example, \cite{lyu97} examined a model
in which the viscosity parameter $\alpha$ varies on the local viscous
time scale. 
Variations which originate at different radii propagate inward to
produce a ``red-noise'' spectrum of variations at the boundary layer.
Although fluctuations in CVs (and CH Cyg) are too fast to be strictly
tied to viscous-time-scale variations in a thin disk, extensive timing
studies of X-ray binaries and active galactic nuclei by
\cite{mcutt01} support this type of 'propagation' model.
As another example, \cite{geerach92} found that in a geometrically
thin, differentially rotating disk, magnetohydro-dynamic turbulent
energy fluctuates at roughly the Keplerian orbital frequency (i.e., on
the dynamical time).  Although the observed flickering will also be
influenced by the dissipation through small turbulent eddies, the
increase in the time scale of fluctuations farther out in the disk
could introduce a radial dependence into the speed of stochastic
variations.

Observationally, numerous authors have associated the dominant source
of 
minute-time-scale flickering in CVs with the inner disk and/or WD
surface
\citep[e.g.,][]{hs85,ofw87,horn94,bru00}.  In several FU Ori
pre-main-sequence stars, which have larger inner-disk radii, the disk
emission varies on time scales closer to a day or less
\citep[e.g.,][]{ken00}.
In neutron-star X-ray binaries, which have smaller inner-disk radii,
stochastic X-ray variations are seen with sub-second time scales
\citep{vdk95}.  In black-hole X-ray binaries, the low-frequency QPO
may shift when the inner-disk radius changes, in the sense that larger
inner-disk radii produce lower-frequency QPOs\footnote{The QPO
frequency in GRO J1655-40, however, has the opposite behavior,
possibly due to relativistic effects \citep[][Varni{\`e}re et
al. 2002]{sob00a}.} \citep[e.g., XTE J1550-564, GRS
1915+105;][]{sob00a,vrt02,mmr99}.
The general connection between proximity to the central object and
variation speed is supported further by the recent discovery that some
QPOs in CVs are analogous to those in low-mass X-ray binaries, but
four orders of magnitude slower
\citep[][]{mauche02}.  Thus, the dependence of variation time scale on location
in the disk is evident both in the comparison among systems with
different size disks and within an accretion disk in a given type of
system (in the sense that slower variations come from the larger disk
radii, and vice versa).

In CH Cygni, the fastest (i.e., minute time scale)
optical variations were absent 
when the 1997 radio jet was ejected; rolling, hour-time-scale
fluctuations remained.
\cite{rod97} reported similar smooth, low-amplitude variations between
1996 June and October, after the drop in optical flux that we have
associated with the collimated jet ejection.
Given the relationship between disk radius and variability time scale
described above, the disappearance of the fastest variations indicates
that the innermost disk was disrupted (or emission from it suppressed)
around the time when the radio jet was produced.
Since \cite{rod97} detected smooth variations as early as 1996 June,
the drop in optical flux just before that time may have been due to
the disappearance of the inner disk.
The reappearance of rapid variations as the flux rose in 1997 August
would then indicate that the inner disk was being rebuilt.

The time scales of both the smooth variations in 1996/97 and the
return of the more rapid fluctuations in late 1997 agree with the
disk-disruption picture.  If, based on the post-jet light curves in
1996/97, we take $\sim 1$ hr as the dynamical time at the inner edge
of the truncated disk, we infer an inner-disk radius after the jet
ejection of $\sim 1\, \rsun$,
\begin{equation} 
R_{inner} \sim 1\;\rsun \left( \frac{t_{dyn}}{1\,hr} \right)^{2/3} \left(
\frac{M_{WD}}{0.5 \msun} \right)^{1/3}.
\end{equation}  
For an accretion rate of $10^{-9} \msyr$ and $\alpha = 0.03$
\citep[for a low-state disk;][]{war95}, $R_{inner} \sim 1\, \rsun$
corresponds to a viscous time of roughly 1 yr,
\begin{equation}
t_{visc} \sim 1\; {\rm yr} \left( \frac{\alpha}{0.03} \right)^{-4/5}
\left( \frac{\mdot}{10^{-9} \msyr} \right)^{-3/10} \left( \frac{R}{\rsun} \right)^{5/4}
\end{equation}  
\citep[see][]{fkr92}.  We expect the disk to be rebuilt on the viscous time
at the inner edge, and in fact the fastest variations return after
roughly 1 yr.

\begin{figure*}
\begin{center}
\epsfig{file=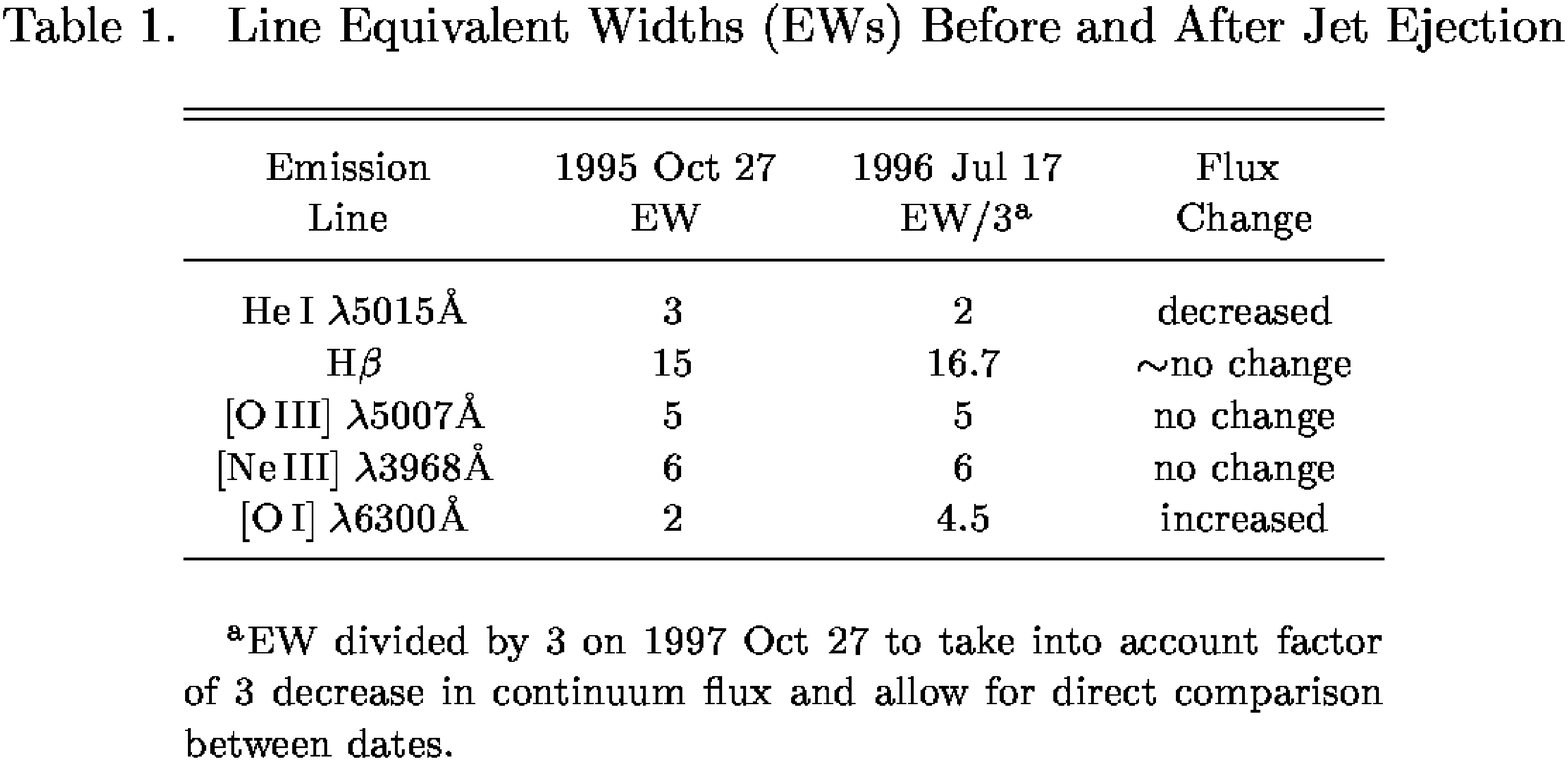,width=5.0in}
\end{center}
\vspace{-1cm}
\end{figure*}

\begin{deluxetable}{cccc}
\tabletypesize{\footnotesize}
\tablecaption{Line Equivalent Widths (EWs) Before and After Jet Ejection\label{tab:postjetews}}
\tablewidth{0pt}
\tablehead{
\colhead{Emission} & \colhead{1995 Oct 27} & \colhead{1996 Jul 17} &
\colhead{Flux} \\
\colhead{Line} & \colhead{EW} & \colhead{EW/3\tablenotemark{a}} &
\colhead{Change}} 
\startdata
He\,I $\lambda$5015\AA & 3 & 2 & decreased \\ H$\beta$ & 15 & 16.7 & $\sim$no change \\ 
$[{\rm O\,III}]\; \lambda$5007\AA & 5 & 5 & no change
\\ $[{\rm Ne\,III}]\; \lambda$3968\AA & 6 & 6 & no change\\ 
$[{\rm O\,I}]\; \lambda$6300\AA & 2 & 4.5 & increased\\
\enddata
\tablenotetext{a}{EW divided by 3 on 1997 Oct 27 to take into account
factor of 3 decrease in continuum flux and allow for direct comparison
between dates.}
\end{deluxetable}

\vspace{-0.3cm}
\begin{inlinefigure}
\plotone{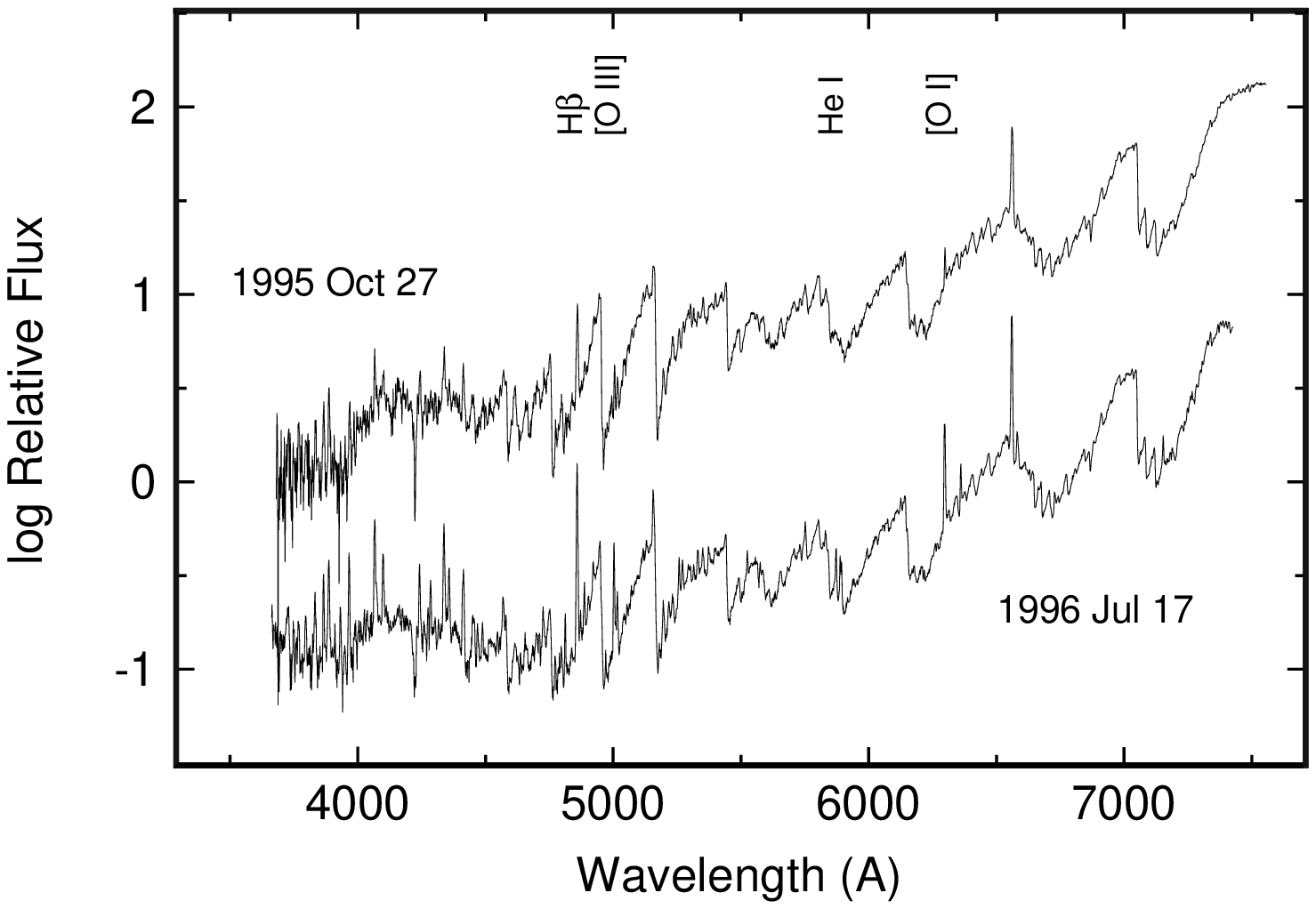}
\caption{
Top: spectrum from 1995 Oct, in the low state before the 1996/1997
radio jet was produced.  Bottom: spectrum from 1996 Jul, after the
drop to the very low state.  The high-excitation He\,I emission-line
flux decreased, whereas the moderate-excitation H$\beta$, [O\,III],
and [Ne\,III] line fluxes remained roughly constant.  [O\,I], which is
a jet line in pre-main-sequence stars,
strengthened. \label{fig:jetspec}}
\end{inlinefigure}

Our velocity extrapolation
indicates that the jet material was expelled during the 1996 flux
drop.  We show example spectra from before and after the optical
decline in Fig.~\ref{fig:jetspec}.  Comparing line equivalent widths,
and taking into account the factor of 3 decline in the continuum (at
$V$), the high-ionization-state He\,I $\lambda$5015 line decreased in
strength after the initial jet ejection, while the
moderate-ionization-state H$\beta$, [O\,III], and [Ne\,III] line
strengths stayed approximately constant.  The [O\,I] line, which is
often associated with jets in pre-main-sequence stars, increased in
strength.
The equivalent-width changes are summarized in
Table~1.
The fact that the He\,I line decreased,
but did not disappear, indicates that 
there may have been a second source of 
photons capable of photo-ionizing He\,I ($\chi = 24.6$ eV),
such as shock-heated colliding winds or the WD surface.  Nevertheless,
the equivalent-width changes are consistent with a decrease in the
ionizing flux from the hot inner disk.

\section{Alternative Interpretations} \label{sec:alt}

We propose that the minute-time-scale variations disappeared in
1997 because the inner accretion disk was disrupted, or emission from
this region decreased, when the radio jet was produced.
There are other alternatives.  In 1997, the fastest variations from
the disk could have been hidden rather than absent.  For example,
opaque material could have blocked the accretion disk.  If the
accreting WD ejected material equatorially before producing the jet
\citep[as in some other jet sources, e.g., SS\,433;][]{para02},
perhaps disk-blocking dust clouds could have formed.
In this picture, the smooth variations would be due to flickering FUV
or soft-X-ray light
that is blocked from direct view, but is seen via reprocessing in the
nebula, which acts as a low-pass filter.
The smooth light curves from late 1996 \citep{rod97} and early 1997
(\S\ref{sec:obs}) are reminiscent of the Balmer line variations in
RS Oph, which presumably originate in the nebula \citep{sokoproc03}.
However, the nebula would need to have a density $n_e \gtrsim 10^9$
cm$^{-3}$ for the H recombination time to be on the order of hours
($n_e \ga 10^{10}$ cm$^{-3}$ for minutes).  Measured nebular densities
for CH Cygni tend to range between one and three orders of magnitude
lower
\citep[e.g.,][]{hack86,mska88}.

Densities in the disk, on the other hand, are much higher.
If, instead of coming directly from the inner disk, the flickering optical
light has been reprocessed in the outer disk, then the change in character
of the light curves in 1997 could be due to a change in the
reprocessing medium.  If the density in the reprocessing site
suddenly decreased, the recombination time would increase, and the
fastest variations would be smeared out.  But then the change we have
observed would still indicate a change in the accretion disk in
conjunction with the production of a radio jet.

Finally, if the accretion is disk-less, 
the frequency content of the flickering could reflect the size
distribution of the blobs of accreted material. 
It is unclear, however, why the size distribution of accreted blobs
would change when a jet is produced.  We therefore favor the
disk-disruption scenario over any of these other models.

\section{Implications for Jet Formation} \label{sec:implications}

We have described observations which indicate that the accretion disk
was involved in jet production in CH Cyg.  In particular, the
disappearance of the fastest flickering implies that material from the
innermost part of the disk was removed, or that emission from this
region decreased.  The removal of material, and the associated energy
and angular momentum, from the inner disk
implies that the jet was accretion-powered.  
Several other authors have proposed that the jets from CH Cygni
originate in an accretion disk, or are the result of ejection of the
inner disk \citep[e.g.,][]{solf87}.

In their study of persistent X-ray binaries, \cite{fh00} found that
similar-strength radio emission, presumably from jets, can exist in
both neutron-star and black-hole systems.  Thus, neither a black hole
nor a compact-star surface is necessary for jet production in these
systems.  On the other hand,
they do not detect radio-jet emission from systems in which the inner
disk is truncated by a strong magnetic field.  Thus, it appears that
the inner accretion disk is involved in, and possibly responsible for,
jet-formation in X-ray binaries. 
Our observations of changes in the accretion disk in CH Cygni indicate
that for CH Cygni also, and therefore possibly for WD jet sources
generally, the inner disk is closely linked to the bi-polar ejection
of material.

In CH Cygni, the production of jets appears to follow a sudden decline in
the optical flux.  The jet-production mechanism, however, is not well
understood. \cite{tay86} proposed that super-Eddington accretion
caused the collimated outflow,
but IUE observations obtained by \cite{mska88} showed that the white
dwarf in CH Cygni was never close to the Eddington limit.  
\cite{mm88} proposed that the jets are driven by rotational
energy from the WD, 
but, as we discuss in paper II, there is little evidence that a
strong surface magnetic field is present.  It is therefore difficult
to extract this rotational energy.
Livio (1997) proposed that WD jets need a source of energy in
addition to disk accretion, such as the nuclear shell burning thought
to exist in supersoft X-ray sources \citep[some of which have
persistent jets;][and references therein]{south97,liv97}, or a hot
boundary layer between the disk and WD surface.  CH Cygni has a low
enough WD luminosity that nuclear burning is unlikely to be a
significant source of energy.  Variations on a time scale as short as
$\sim100$ s in the hard X-rays
\citep{ezu98,lt87}, as well as the high-ionization-state optical and
UV
emission lines suggest that it may, however, have a boundary layer.

The disk in CH Cygni is probably more similar to CV disks than disks
that involve an advection-dominated accretion flow (ADAF), or other
structures that have been suggested for black-hole systems
\citep[such as advection-dominated inflow-outflow solutions (ADIOS),
or ``sphere + disk'' Comptonization
models;][]{ny95,esin98,bb99,wilms99}.  Nevertheless, our observations
of the flickering in CH Cygni show that the inner radius appears to
change when a jet is produced, as seen in some X-ray binaries.  X-ray
spectral fitting of the black-hole X-ray transient XTE J1550-564
indicates that the disk inner radius increased at the beginning and end
of an X-ray outburst, at the same time that material was ejected
\citep{sob00b,fk01}.  In the black-hole candidate GRS 1915+105, the
inner disk was repeatedly evacuated when material was ejected
\citep{bell97a,bell97b,fer99}. 

\cite{bell97a} suggest that a dwarf-nova-like disk instability
may be responsible for the observed limit cycle behavior and
associated mass ejections of GRS 1915+105.  In pre-main-sequence stars
as well, the strength of the collimated outflow appears to be
associated with the state of a thermally unstable accretion disk
\citep{hk96}.  In paper II, we propose that the activity in CH Cygni
may also be driven by an unstable disk.  Thus, although the inner disk
around black holes and pre-main-sequence stars may be quite different
from the inner disks around accreting WDs,
the physics of jet ejection provides a link between them.

If discrete ejections are associated with state transitions in an
unstable disk, it is still not clear whether disk thermal
instabilities (DTIs) lead to collimated ejections or vice versa.
\cite{fh00} suggest that in X-ray binaries, the ejections are a
by-product of an extreme physical change in the accretion flow.  We
mention one possible case in which a jet
\citep[say, due a centrifugally driven wind, such as in
the][bead-on-a-string model]{bp82} leads to a DTI.
Knigge (1999)
showed that the disk temperature profile is modified with the addition
of either a radiation-driven or centrifugally driven wind.
In either case, 
the central temperature in a disk+wind system is lower, sometimes
significantly, than a disk alone.  Theoretical calculations of CV disk 
winds have shown that the mass-loss rate in the wind increases with
disk luminosity for a bright disk \citep{proga99}, so a high-state
disk produces a strong wind.  The presence of a strong enough
disk wind from the inner disk, or a centrifugally driven jet, could
therefore potentially trigger an inside-out cooling wave.

In summary, there are some intriguing observational similarities
between CH Cygni and X-ray binaries with neutron-star or black-hole
accretors.  Both CH Cygni and X-ray transients can eject hot plasma
when the systems change state.  In CH Cygni, the jet ejections are
associated with the sudden drop from an optical high state to an
optical low state.  In this paper, we have described a second similarity --
changes in CH Cygni's light curve which suggest that emission from the
inner disk decreased substantially when a jet was produced,
possibly due to the evacuation of this region.  These observational
similarities support the idea that jets can form in the same way in
different types of systems.

\acknowledgments

This work has been supported in part by NSF grant INT-9902665.  In
this research, we have used and acknowledge with thanks, data from the
AAVSO International Database, based on observations submitted to the
AAVSO by variable star observers worldwide.  We would like to thank C.
Brocksopp, M. Muno, and L. Bildsten for useful comments and
discussion.

\clearpage

\end{document}